\documentclass[journal]{IEEEtran}

\usepackage[utf8]{inputenc}
\usepackage[T1]{fontenc}
\usepackage{graphicx} 
\usepackage{amsmath}
\usepackage{amsfonts}
\usepackage{caption}
\usepackage{float} 
\usepackage{cite}
\usepackage{subcaption}
\usepackage{color}
\usepackage{lineno}
\usepackage{amssymb}

\usepackage{booktabs}
\usepackage{multirow}

\ifCLASSINFOpdf
  
\else
  
\fi

\begin{document}

\title{\LARGE \bf
    Temporal Information Reconstruction and Non-Aligned Residual in Spiking Neural Networks for Speech Classification 
}
\author{Qi~Zhang,~Huamin~Wang,~Hangchi~Shen,~Shukai~Duan
    ~\IEEEmembership{Member,~IEEE,}
        Shiping~Wen,~\IEEEmembership{Senior Member,~IEEE,}%
        and~Tingwen~Huang,~\IEEEmembership{Fellow,~IEEE}
 \thanks{Manuscript received; revised; accepted. Date of publication ; date of current version. This work was supported in part by Fundamental Research Funds for the Central Universities under Grant SWU021002, in part by the Project of Science and Technology Research Program of Chongqing Education Commission under Grant
 KJZD-K202100203, and in part by National Natural Science Foundation of
China under Grant U1804158. (Corresponding author: Huamin Wang.)}
\thanks{Qi Zhang, Huamin Wang, Hangchi Shen and Shukai Duan are with College of Artificial Intelligence, Southwest University, Chongqing, 400715, China, and also with Chongqing Key Laboratory of Brain Inspired Computing and Intelligent Chips, Chongqing, 400715, China (e-mail:myhonor151@email.swu.edu.cn; hmwang@swu.edu.cn; stephen1998@email.swu.edu.cn; duansk@swu.edu.cn)}
\thanks{Shiping Wen is with College of Information Engineering, University of Technology Sydney, Australian Institute of Artificial Intelligence, Sydney, 2007, Australia (e-mail: 
Shiping.Wen@uts.edu.au).}
\thanks{Tingwen Huang is with Faculty of Computer Science and Control Engineering, Shenzhen University of Advanced Technology, Shenzhen 518055, China (e-mail: huangtw2024@163.com).}}

\markboth{}%
{Shell \MakeLowercase{\textit{et al.}}: Bare Demo of IEEEtran.cls for IEEE Journals}

\maketitle

\begin{abstract}
Recently, it can be noticed that most models based on spiking neural networks (SNNs) only use a same level temporal resolution to deal with speech classification problems, which makes these models cannot learn the information of input data at different temporal scales. 
Additionally, owing to the different time lengths of the data before and after the sub-modules of many models, the effective residual connections cannot be applied to optimize the training processes of these models.
To solve these problems, on the one hand, we reconstruct the temporal dimension of the audio spectrum to propose a novel method named as Temporal Reconstruction (TR) by referring the hierarchical processing process of the human brain for understanding speech. Then, the reconstructed SNN model with TR can learn the information of input data at different temporal scales and model more comprehensive semantic information from audio data because it enables the networks to learn the information of input data at different temporal resolutions. 
 On the other hand, we propose the Non-Aligned Residual (NAR) method by analyzing the audio data, which allows the residual connection can be used in two audio data with different time lengths.
 We have conducted plentiful experiments on the Spiking Speech Commands (SSC), the Spiking Heidelberg Digits (SHD), and the Google Speech Commands v0.02 (GSC) datasets. According to the experiment results, we have achieved the state-of-the-art (SOTA) result 81.02\% on SSC for the test classification accuracy of all SNN models, and we have obtained the SOTA result 96.04\% on SHD for the classification accuracy of all models. Furthermore, on the non-spiking dataset GSC, we can achieve a test classification accuracy of 95.63\%, which surpasses the original baseline SNN-Delays.
 Meanwhile, the method proposed in this paper can speed up the model's processing speed and reduce its energy consumption.
\end{abstract}

\IEEEpeerreviewmaketitle

\section{Introduction}

\IEEEPARstart{P}{rofoundly} inspired by biological systems\cite{doi:10.1142/S0129065709002002}\cite{Xu_2023_CVPR}, Maass provided the spiking neural networks (SNNs), named as the third-generation artificial neural networks (ANNs)\cite{MAASS19971659}, which can meticulously mimic the operational mechanisms of biological neural systems\cite{9597475}. 
Compared to ANNs, SNNs can achieve a transformation from static to dynamic in information representation, computing patterns, and learning mechanisms\cite{10003253} by using the timing and the frequency of spike releases to encode information, which allows the state of neurons to evolve dynamically over time\cite{wu2022brain,10168967}. 
Furthermore, spikes can be triggered only when the membrane potential of the neurons exceeds the threshold\cite{Liao_Liu_Zheng_Pan_2024,bu2023optimal}. This means that SNNs exhibit a dynamic discrete characteristic, and only have sparse accumulation (AC) instead of costly multiply-accumulate (MAC) operations\cite{pmlr-v202-wang23j,shen2024rethinking}, thereby demonstrating higher energy efficiency\cite{9540752,Wang_Zhang_Han_Wang_Zhang_Xu_2023,10254579}. Therefore, SNNs have been applied in many fields to reduce the energy consumption in recent years, such as pattern recognition\cite{MA2025106765,10032591}, image classification\cite{ma2025neuromoco,9646435}, speech classification\cite{stcc-SNN,10.3389/fnins.2023.1275944,2023arXiv230617670H}, and etc. 

Due to the dynamic discrete characteristics of SNNs, they can efficiently capture the dynamic changes and fully leverage the temporal information, making them particularly suitable for handling complex data with temporal dependencies, such as speech data. As a result, when SNNs are used for speech classification, they can demonstrate some significant advantages.
Cramer et al.\cite{9311226} introduced the Spiking Heidelberg Digits (SHD) dataset and trained a single-layer spiking recurrent neural network (SRNN)\cite{10.1162/neco.2009.11-08-901} on it for the first time. Base on SRNN, the authors in \cite{yin2021accurate} provided adaptive spike neurons to dynamically adjust their firing rate and threshold according to the changes of the input current and achieve high performance. 
Then, Bittar et al.\cite{10.3389/fnins.2022.865897} introduced an additional adaptive variable $w(t)$ that linearly coupling with the membrane potential $u(t)$ to reproduce various special firing patterns observed in the biological neurons. 
Dampfhoffer et al. constructed the SpikGRU\cite{10.1007/978-3-031-15934-3_30} model by adding an additional gate to the cuba-LIF\cite{gerstner2002spiking} to calculate the current state through the optimal combination of the previous state and input current, which could achieve higher accuracy than other SRNNs and reduce a large number of operations compared to ANNs.

In recent years, with the proposition of attention mechanisms, they have gradually been incorporated into SNNs to process speech data, resulting in more significant effects\cite{stcc-SNN}\cite{Yao_2021_ICCV}.
In \cite{Yao_2021_ICCV}, the attention concept was extended to the temporal dimension to discard the irrelevant frames, name as TA-SNN that can extract effective spatio-temporal features from event streams and consider the changes in frame signal-to-noise ratio.
After that, STSC-SNN\cite{stcc-SNN}incorporated attention mechanism and temporal convolution to achieve the filtering and gating functions of synapses, endowing the synapses with temporal dependence and enhancing their ability to classify speech information.
As we all know, connection delay among spiking neurons may affect the performance of models, which makes some authors to investigate delay problem. 
Sun et al.\cite{10.3389/fnins.2023.1275944} designed an adaptive training scheduling mechanism allowing the axon delays to be learned, which implied that it could be adapted to the learning of each layer in the SNNs to improve the performance of speech classification tasks.
Based on dilated convolution with learnable spacings (DCLS), Hammouamri et al.\cite{2023arXiv230617670H} built the SNN-Delays to learn the connection delay between neurons, achieving the SOTA results on multiple datasets.

Additionally, the process of speech comprehension is generally viewed as a series of computational steps that are carried out by  different processing modules with distinct functional role in the brain, each of which is used to extract information on different temporal scales\cite{BORNKESSELSCHLESEWSKY2015142}\cite{deHeer6539}. Therefore, learning multi-temporal information from speech signals has great value and significance in speech classification, and has attracted some researchers' attention\cite{8047110}. Multiscale audio spectrogram transformer (MAST)\cite{10095023} introduced the concept of multi-scale feature hierarchy into audio spectrogram transformer (AST), allowing MAST to learn information at different temporal resolutions and achieving excellent results in speech classification tasks. Moreover, according to the published research results, it has found that SNN models possess spatial-temporal characteristics, giving them an advantage in handling speech classification tasks. However, as far as we know, there have few works that use SNN models to deal with the information of multiple temporal scales for speech classification problems.

Motivated by the above analysis, we have proposed the Temporal Reconstruction (TR) method to achieve multi-time-scale modeling of input speech information in speech classification tasks by referencing the hierarchical processing of speech information in the human brain. In addition, we have also proposed the Non-Aligned Residual (NAR) method to enable residual connections to be used on speech data with different lengths in the temporal dimension.
We have conducted a great deal of experiments on the spiking datasets SSC and SHD to validate the effectiveness of our proposed methods. And we have also conducted some experiments on the non-spiking dataset GSC to further validate the effectiveness of our methods. The contributions of this paper can be summarized as follows:

\quad (1) In analogy with the hierarchical processing of speech information in the human brain, a novel TR method has been proposed to reconstruct the temporal dimension of the audio spectrum, which enables the network to learn information at different time scales of the input audio spectrogram.

\quad (2) A NAR method has been proposed to establish residual connections between speech data with different time lengths, which can solve the problem that the residual connections cannot be directly used to process speech data of different time lengths.
    
\quad (3) Promising experimental results have been obtained on the spiking datasets SSC and SHD, and the non-spiking dataset GSC by a great deal of experiments using our proposed methods. On the dataset SSC, a SOTA result of 81.02\% has been achieved for the test classification accuracy among all SNN models, and a SOTA result of 96.04\% has been obtained on the dataset SHD for the test classification accuracy across all models. At the same time, the methods proposed in this paper can also achieve a favorable energy efficiency ratio on the non-spiking dataset GSC.

The structure of this paper is as follows. In Section \ref{sec:related_work}, we introduce the spiking neuron model known as the leaky integrate-and-fire (LIF) model, the learning mechanism and components of the SNN-Delays model, related work on multi-scale information learning, and research on residual connections. In Section \ref{sec:methods}, we describe the detailed implementation of our proposed TR and NAR. In Section \ref{sec:experiments}, we present the experimental results of our methods and the results of ablation experiments. In Section \ref{sec:conclusion}, we conclude our paper.

\section{Related works}
\label{sec:related_work}

\subsection{Spiking Neuron Model}
\begin{figure}[htbp]
    \centering
    \includegraphics[width=1\columnwidth]{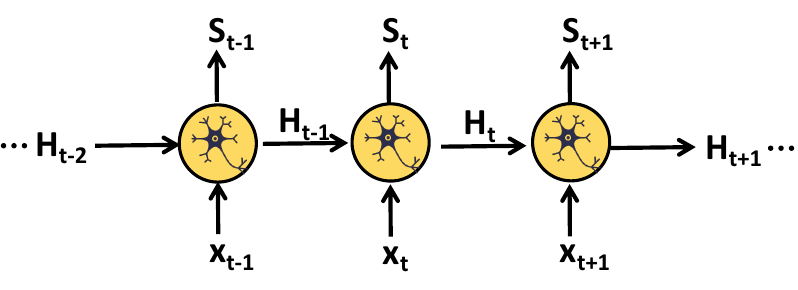} 
    \caption{LIF neurons}
    \label{fig-lif}
\end{figure}
There are several neuron models available in SNNs, including the integrate-and-fire (IF) model \cite{burkitt2006review}, the LIF model \cite{gerstner2002spiking}, and others. The LIF model is widely used due to its simplicity and efficiency. In an LIF neuron, a spike is emitted only when the membrane potential exceeds the threshold. After the spike is fired, the membrane potential is reset, and there is a degree of attenuation of the membrane potential at every moment. In our work, we choose to employ the LIF model and implemented by spikingJelly\cite{doi:10.1126/sciadv.adi1480}. The LIF model is shown in Figure \ref{fig-lif} and is described by the following equations:
\begin{equation}
\label{eq-lif1}
\begin{cases}
{I}[t]=W \cdot X[t]+b,\\U[t]=\eta \cdot H[t-1]+I[t],\\S[t]=Heav(U[t]-V_{th}),
\end{cases}
\end{equation} 
\begin{equation}
\label{eq-lif2}
H[t]=\begin{cases}U[t]\cdot(1-S[t])+V_{reset}\cdot S[t],&\text{hard reset},\\U[t]-V_{\mathrm{th}}\cdot S[t],&\text{soft reset},\end{cases}
\end{equation}
where $X[t]$ represents the input to the LIF neuron model at time $t$, with $W$ and $b$ denoting the synaptic weights and bias, respectively. $I[t]$ denotes the input current at time t, while $U[t]$ is the voltage at that moment. The parameter $\eta$ is the decay constant, and $H[t-1]$ represents the voltage value after reset at time $t - 1$. $S[t]$ indicates the firing activity at time $t$, while $Heav(*)$ refers to the Heaviside function. $V_{th}$ is the firing threshold, and $V_{reset}$ is the reset voltage.
\subsection{Baseline Architecture}
In \cite{2023arXiv230617670H}, the author proved the mathematical equivalence between 1D temporal convolution and connection delay, built the SNN-Delays network using the dilated convolution with learnable spacings (1D version) proposed by Khalfaoui-Hassani\cite{khalfaouihassani:hal-04057309}, and achieved the SOTA on three audio classification datasets: SHD, SSC, and GSC. Its core innovation lies in learning both synaptic weights and connection delays between neurons. The network consists of multiple delay modules, each of which contains a DCLS1D, Batch Normalization, LIF Neuron, and Dropout in sequence, with no Batch Normalization and Dropout in the output layer.
The connection delay refers to the varying time taken when a spike signal passes through different structural synapses between two neurons. Maass\cite{MAASS199926} has proved that the delayed learning is important in theory and the learning in the brain cannot be simplified to synaptic plasticity. Based on these, we have considered SNN-Delays of \cite{2023arXiv230617670H} as the baseline in this paper, and have successfully achieved SOTA on both the SSC and SHD datasets by combining our methods.

\subsection{Multi-scale Information Learning}
In the field of natural language processing and visual fields, multi-scale information learning has been used in some published works\cite{10095023,Liu_2021_ICCV,9746312}.
In \cite{10095023}, Ghosh et al. built the MAST by introducing the concept of multi-scale feature hierarchy, and achieved significantly higher classification accuracy on LAPE's 8 tasks than AST. Meanwhile, MAST could also reduce the square attention complexity in AST due to the lower time resolution of the input data. As we know, MAST can process data with high temporal resolution at the shallow layers, which means that it is focused on local information of the data and is dedicated to modeling simple low-level acoustic attributes. While at the deep layers, it can process data with low temporal resolution, which means that it is focused on global information of the data and learns complex high-level acoustic attributes. Motivated by this method, we propose TR to enable the SNNs to learn information at different temporal scales from the input data for improving speech classification, which is shown in Section \ref{sec:methods}.

\subsection{Residual Connections}
In 2016, the concept of residual connections was first introduced by Kaiming He et al. \cite{He_2016_CVPR}. They not only provided strong theoretical support but also validated its significant advantages in image recognition by leveraging the additive properties of signals. In the field of SNNs, some works have involved residual connections to improve the models' performance\cite{SHEN2024120136,10032591,10486935}.
Given this, we believe that applying residual connections to our baseline will also yield promising results. However, due to the special padding method of SNN-Delays, the lengths of the two data that require residual connections differ in the temporal dimension. To address this, we propose NAR that is described in the section \ref{sec:methods}.

\section{Methods}
\label{sec:methods}
\begin{figure*}[htbp]
    \centering
    \makebox[\textwidth][c]{\includegraphics[width=1\textwidth]{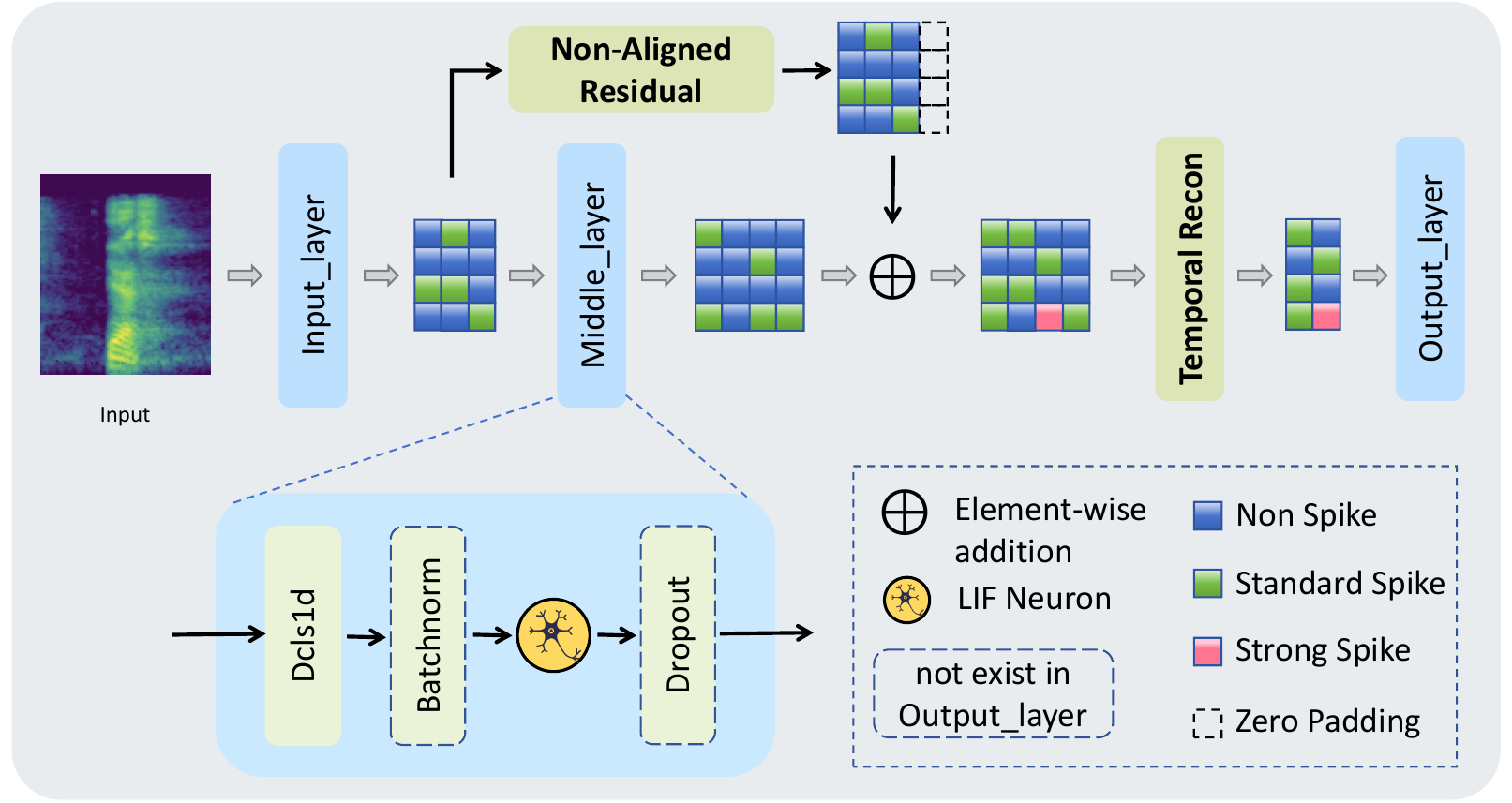}}%
    \captionsetup{justification=raggedright, singlelinecheck=false} 
    \caption{The structure of the SNN-Delays model that incorporates Temporal Reconstruction (TR) and Non-Aligned Residual (NAR) methods. TR allows the model to learn information carried by multiple timescales in the data, and NAR enables residual connections to be applied to the data of varying time lengths.}
    \label{fig-jiagou}
\end{figure*}
In the following two subchapters, we will provide a detailed description of our proposed TR and NAR methods, which is shown in Figure \ref{fig-jiagou}.
\subsection{Temporal Reconstruction (TR)}
The core strategy of TR is to first group the data along the temporal dimension, and the number of time points in each group is named as within-group time length. Subsequently, within each group, we perform comparative analysis in the frequency dimension for each discrete time point, and reconstruct multiple time points into a single time point. Specifically, in a group, we can obtain the value of a frequency point at the new reconstruction time by logic OR operation along the temporal dimension at that frequency point. Then, we repeat the above operation along the frequency dimension to complete the TR. If a residual connection is used before TR, the Max function can be used instead of the logical OR operation. 
 
TR method can learn information at different temporal scales from the input audio spectrogram by reducing its temporal resolution. At shallow layers, it allows the model to learn local information and extract low-level acoustic attributes such as loudness, pitch, and timbre. While at deep layers, it can study global information and extract high-level acoustic attributes such as dynamic range, audio quality defects, and emotional attributes. Furthermore, as the length of the data in the temporal dimension decreases, the processing speed of the model will become faster. We have two versions of TR: TR with overlap and TR without overlap, based on whether the time points between groups overlap.


Based on whether the chosen group-within time length can be evenly divided by the total time length of the input data, TR-o is divided into two cases: one that is evenly divided and the other that isn't.
The implementation formula of TR-o is shown by the following equation:
\begin{equation}
\label{eq-o}
    \begin{cases}
    \underset{[0:T_1:1]}{\tilde{x}^{\tilde{t}}} = \underset{[0:T_2:stride]}{Max}(x^i,\ldots,x^{i+len-1}),\hspace{0.2cm}\text{if } T \% T_{1} \, = \, 0,\\
    \underset{[0:T_1:1]}{\tilde{x}^{\tilde{t}}} = \underset{[0:T_2:\tilde{stride}]}{Max}(x^i,\ldots,x^{i+len-1}) \oplus x^{t_{last}} ,\hspace{0.2cm} \text{else},
    \end{cases}
\end{equation}
where $\tilde{x}$ represents the data after TR, and $x$ represents the original data, $x$ and $\tilde{x}$ belong to $\mathbb{R}^{(T,B,N)}$, $T$ represents the size of the temporal dimension, $B$ represents the batch size, $N$ represents the size of the frequency dimension, stride is the step size relative to time points, len is the within-group time length, $T_1=\lfloor \frac{T-1}{stride}\rfloor$, $T_2=(\lfloor \frac{T-1}{stride} \rfloor-1)*stride$, and the symbol "$\oplus$" represents the concatenate operation in temporal dimension. 

\begin{figure}[htbp]
    \centering
    \includegraphics[width=0.8\columnwidth]{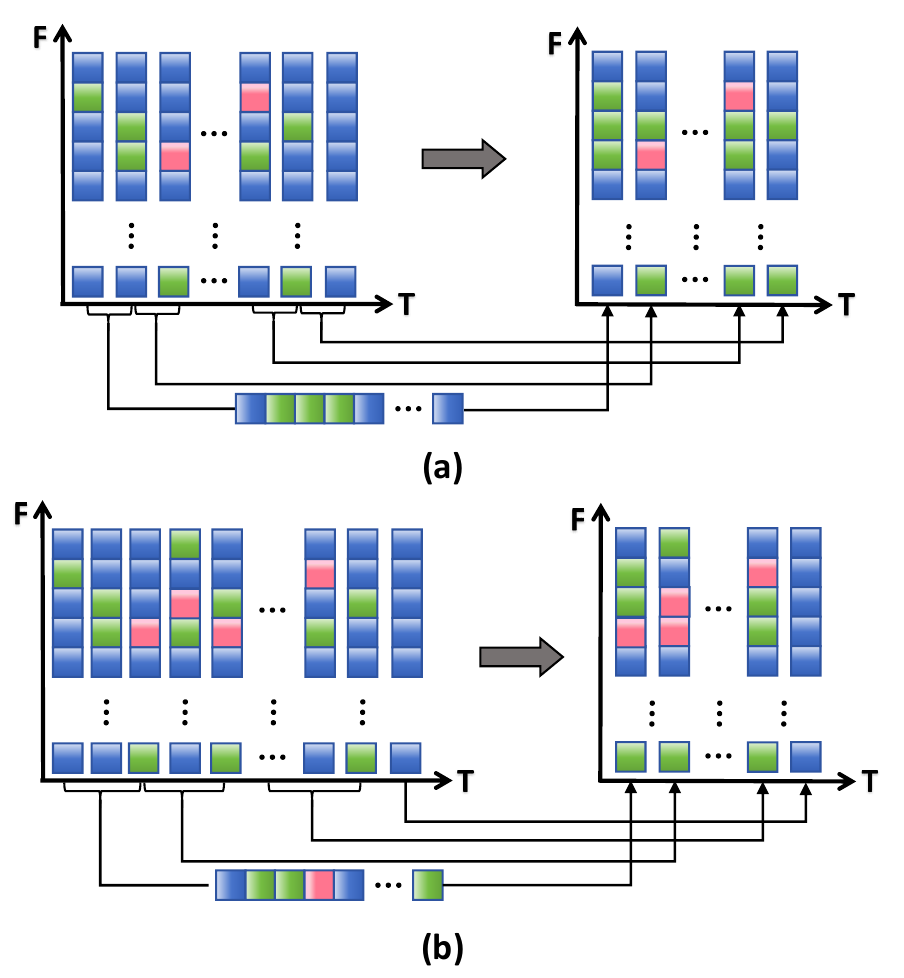} 
    \captionsetup{justification=raggedright, singlelinecheck=false} 
    \caption{\small \small TR-o (TR with overlap). The blue blocks represent no spikes, the green blocks represent standard spikes, and the pink blocks represent strong spikes. (a) A schematic diagram when the group-within time length can be evenly divided by the total time length with group-within time length is 2 and stride is 1. (b) A schematic diagram when the group-within time length cannot be evenly divided by the total time length with group-within time length is 3 and stride is 2.}
    \label{fig-tro}
\end{figure}

The schematic diagram of the TR-o method is shown in Figure \ref{fig-tro}.
When the chosen group-within time length can be evenly divided by the total time length of the input data, as shown in Figure \ref{fig-tro} (a). We assume that the shape of the data before TR-o is $(t_1,f)$, where $t_1$ represents the temporal dimension and $f$ represents the frequency dimension. The green square in the figure represents the presence of spike, the blue square represents the absence of spike and the pink square represents the presence of strong spike. The within-group time length is set to 2, and the stride is set to 1. In this setting, the specific implementation of TR-o is as follows: First, the first and second time points are grouped into one group, scanning along the frequency dimension. If a frequency point appears at least one strong spike in both time points, the frequency point at the new reconstruction time point is marked as having a strong spike. If there are no strong spikes at either time point, we check for the presence of any spikes at this two time points. If spikes are present, the frequency point is marked as a spike at the new reconstruction time; otherwise, it is marked as having no spike. Subsequently, the second time point and the third time point, the third time point and the fourth time point, etc. Finally the processed data shape is $(t_1-1,f)$. When the chosen group-within time length can not be evenly divided by the total time length of the input data, as shown in Figure \ref{fig-tro} (b). We assume that the shape of the data before TR-o is $(t_2,f)$, with a within-group time length is 3 and a stride is 2. Unlike the previous case, the length of the last group is less than the selected within-group time length; therefore, we directly choose the data of the last time point as the output of the last group and then remove the data of the last group , process the remaining data as shown in Figure \ref{fig-tro} (a), and finally concatenate the processed data with the data from the last time point along the temporal dimension. In the end, we obtain the shape of processed data is $(\lfloor \frac{t_2-1}{3}\rfloor+1,f)$.

Similar to TR-o, TR-no is also divided into two cases: one that is group-within time length can evenly divided by total time length and the other that isn't. The implementation formula for TR-no is shown by the following equation:
\begin{equation}
\label{eq-no}
    \begin{cases}
    \underset{[0:T_3:1]}{\tilde{x}^{\tilde{t}}} = \underset{[0:T4:\tilde{len}]}{Max}(x^i,\ldots,x^{i+len-1})  ,\hspace{0.2cm}\text{if \ T\%len\,=\,0},\\
    \underset{[0:T_3:1]}{\tilde{x}^{\tilde{t}}} = \underset{[0:T_4:\tilde{len}]}{Max}(x^i,\ldots,x^{i+len-1}) \oplus x^{t_{last}} ,\hspace{0.2cm} \text{else,}
    \end{cases}
\end{equation}
where $T_3=\lfloor \frac{T}{len+stride} \rfloor$, $\tilde{len}=len+stride$, $T_4 = (\lfloor \frac{T}{len+stride} \rfloor-1)*\tilde{len}$, and the stride here refers to the relative time step between each group. 

\begin{figure}[htbp]
    \centering
    \includegraphics[width=0.8\columnwidth]{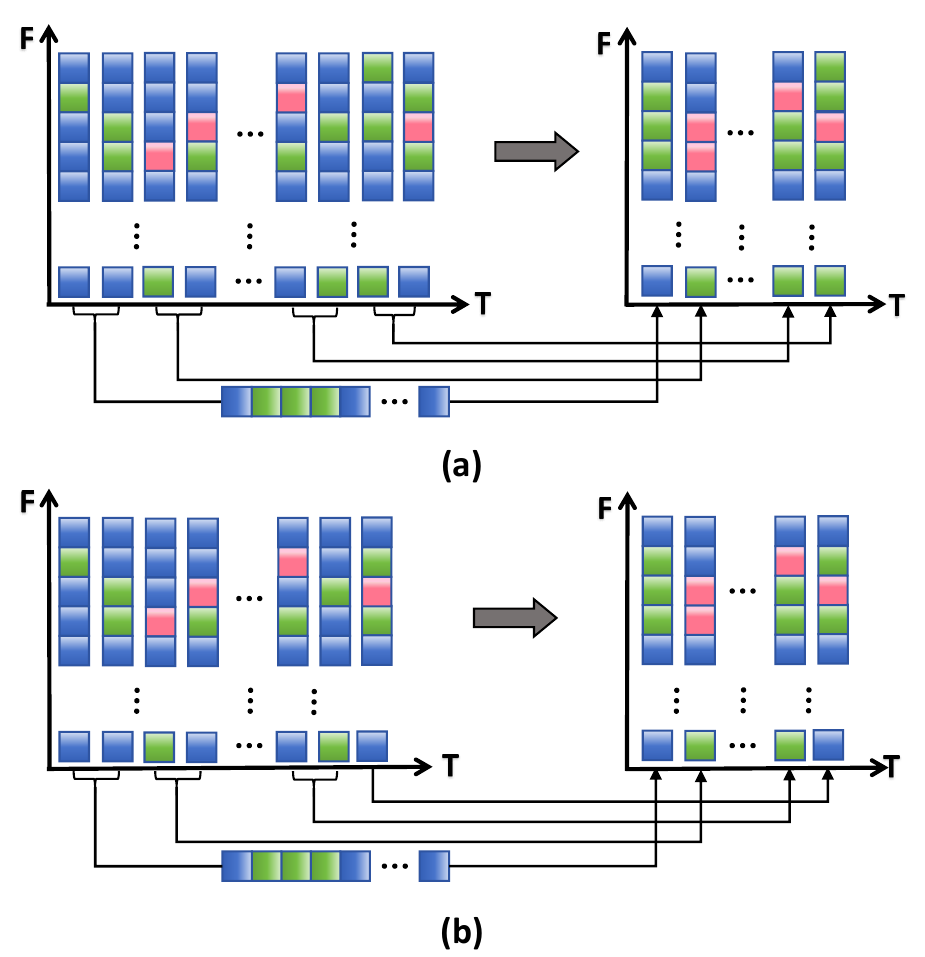} 
    \captionsetup{justification=raggedright, singlelinecheck=false} 
    \caption{\small \small TR-no (TR without overlap). The meanings represented by the blocks of different colors are the same as those in Figure \ref{fig-tro}. (a) A schematic diagram when the group-within time length can be evenly divided by the total time length. (b) A schematic diagram when the group-within time length cannot be evenly divided by the total time length.}
    \label{fig-trno}
\end{figure}

When the group-within time length can be evenly divided by the total time length of the input data, as shown in (a) in the Figure \ref{fig-trno}, the data shape is assumed to be $(t_3,f)$, the group-within time length is 2, and the stride is 1. Under this settings, the frequency information at the first time point and the second time point are processed using an logic OR operation, followed by the processing of the information at the third and fourth time points, and so on, until the processing of the information at the final two time points is complete. The processed data has a shape of $(t_3/2,f)$. When the group-within time length cannot be divided evenly by the total time length, the basic steps are the same as when it can be divided evenly, with the difference being that the last group is first processed, whose time length is less than the group-within time length, directly select the frequency information of the last time point as the output of the last group and then delete the data of the last group from the total data. Finally, the data of the last time point and the data of other groups that have been processed are concatenated in the temporal dimension. As shown in (b) in the Figure \ref{fig-trno}, the last group has only one time point's frequency information, which is the output of the last group. When we assume the data shape to be $(t_4,f)$, the final output after TR-no has a shape of $(t_4/2 + 1,f)$.
\subsection{Non-Aligned Residual (NAR)}
To overcome the problem that residual connections cannot be directly used for SNN-Delays model, we have designed the NAR method to allow residual connections to be flexibly applied in this model, thereby effectively enhancing its information processing ability.
Since the padding method of SNN-Delays results in different time lengths of the data before and after a delay module, the time lengths of the two data need to be processed to the same length in order to apply the residual connections to the model. After the residual connection, the local data will appear values 0, 1, and 2, where 2 can be considered as strong spike improving the expression ability .

Through a lot of experiments, we found that using 0 for right padding is the best approach. On account of the property of learning delay for the SNN-Delays model, left padding will destroy the connection delay that the model has learned. Additionally, in the field of linguistics and phonetics, it is mentioned that the volume and pitch of speech may gradually decrease, especially at the end of a sentence, pause, or the end of a speech, which is called intonation decline or the diminution of the volume, so it is more reasonable to use 0 to padding the data in this method\cite{ladd2008intonational}.

\begin{figure*}[htbp]
 \centering
 \makebox[\textwidth][c]{\includegraphics[width=1\textwidth]{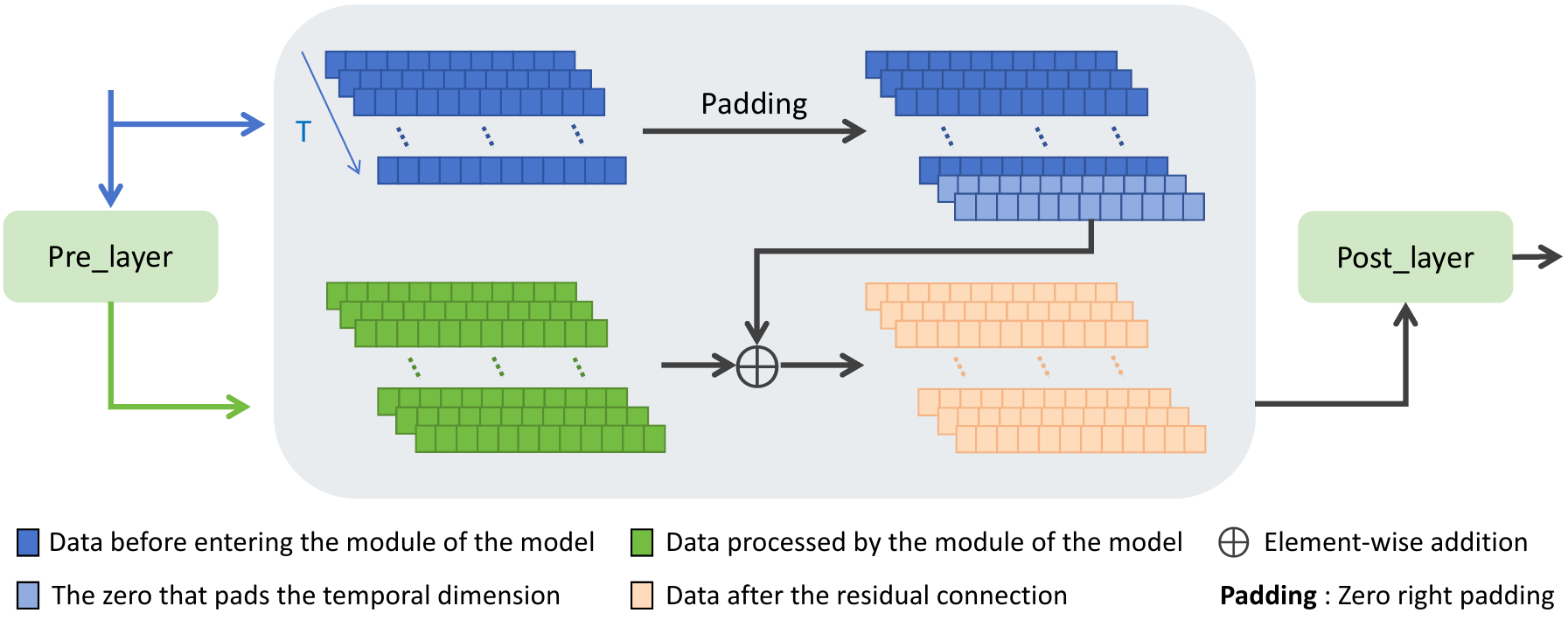}}%
 \captionsetup{justification=raggedright, singlelinecheck=false} 
 \caption{\small \small NAR. This method can allow residual connections to be applied to the data of varying time lengths.}
 \label{fig-rc}
\end{figure*}

The specific operation process is shown in the Figure \ref{fig-rc}. First, the difference between the two data in the temporal dimension is calculated, and then the shorter data in the temporal dimension is padded with 0 on the right to supplement it. The length of the padding is the difference between the two data in the temporal dimension. After padding, the two data can be residual connected, as shown in the following formula:
\begin{equation}
\label{eq1}
y=D(x)+NAR(x,D(x)),
\end{equation}
where $D(*)$ represents a module, $NAR(*)$ represents NAR, and the return value of $NAR(*)$ is $x$ after padding, and the shape of $NAR(x,D(x))$ is exactly the same as that of $D(x)$. When NAR is applied to the model, the performance of the model improves significantly, and the specific experimental data can be found in \ref{sec:experiments} of this paper.

\section{Experiments}
\label{sec:experiments}
In this section, we have conducted numerous experiments on the datasets SSC, SHD, and GSC to verify the effectiveness and advancement of our proposed methods. 
\subsection{Datasets}
\textbf{SHD.}The Spiking Heidelberg Digits (SHD) dataset is a classification dataset based on audio, which is derived from the Heidelberg Digits (HD) dataset through spike conversion. SHD contains approximately 10k high-quality English and German spoken digit recordings ranging from 0 to 9. There are 12 speakers, including 6 women and 6 men. Speakers ranged in age from 21 to 56, with an average age of 29.

\textbf{GSC.}The Google Speech Commands (GSC) dataset is released under the Creative Commons BY 4.0 license and contains the words spoken by 1864 speakers. The GSC contains a total of 35 categories, which include 24 word commands such as Yes, No, Up, Down, Left, Right, On, and Off, as well as 10 auxiliary words such as Bed, Bird, Cat, Dog, and silence.

\textbf{SSC.}The Spiking Speech Commands (SSC) dataset is converted from GSC using an artificial model of the inner ear and parts of the ascending auditory pathway\cite{9311226}.

\subsection{Experiments Setup}
For the basic hyperparameters of the models, we generally follow the settings of the original baselines. We have applied TR and NAR behind the middle layer of baselines, and have achieved the best results by using SNN-Delays as the baseline. The middle layer of SNN-Delays consists of one or more delay modules, and its structure has been described in Section \ref{sec:related_work} and \ref{sec:methods}. 
For the SHD dataset, since its middle layer only has one delay module, we first perform NAR between the input and output of the middle layer. Next, we apply TR-no on the residual connection result, selecting the within-group time length to be 3 and the stride to be 1, and then send the result to the output block. 
For the SSC and GSC datasets, since the middle layer of the model is composed of 2 delay modules, we use NAR behind each delay module in middle layer. Then, if we use TR-no, we set the within-group time length to 3 or 2 with a stride of 1. For TR-o, we set the within-group time length to 2 and the stride to 1.

\subsection{Throughput}
Throughput refers to the number of samples that the model can process in one second. We use the shape of the first batch of samples from each dataset as the standard batch size for calculating throughput, and then generate a tensor of the corresponding size with batch size using PyTorch. By inputting this tensor into the respective model and running it one thousand times, we can calculate the number of samples processed by the model in one second, which is the throughput of this paper.
\subsection{Energy Consumption}
The energy consumption for synaptic computation mainly comes from the accumulator $(0.9 pJ)$ and the multiplier $(4.6 pJ)$ in the neuromorphic chip\cite{6757323}. For pure SNN models, the energy consumption is calculated by first determining the number of activated synapses and then multiplying that number by the energy consumption of the accumulator per operation. If the model includes ANN components, it is necessary to calculate the $FLOPS$ for those components, the energy consumption for the ANN part is then the corresponding $FLOPS$ multiplied by the energy consumption of the multiplier per operation. Therefore, the formulas for calculating energy consumption are as follows:
\begin{equation}
\label{eq-pec}
EC_{pure}=Synapsed^{SNN}_{activated} \times 0.9
\end{equation}
\begin{equation}
\label{eq-npec}
\begin{split}
EC_{non-pure} & = Synapsed^{SNN}_{activated} \times 0.9 \\
              & \quad + FLOPS^{ANN} \times 4.6
\end{split}
\end{equation}
\subsection{Result}

Firstly, by using the SNN-Delays model as the baseline, we have validated the advancement and effectiveness of our methods on the SSC dataset, as is shown in Table \ref{tab-ssc}. From Table \ref{tab-ssc}, it can be found that throughput, energy consumption, and classification accuracy have been improved by our methods.
Especially, if the SNN-Delays (3L-2KC) is integrate with the TR and NAR method, 
the classification accuracy on the SSC dataset has been improved by 0.33\% to reach 81.02\%, establishing a SOTA result for all SNN models on the SSC dataset.

\begin{table*}[htbp]
\caption{Classification accuracy, parameter count, throughput \\ and energy consumption of different models on SSC}
\label{tab-ssc}  
    \centering
	\begin{tabular}{cccc cc}
		\toprule	
	    &Methods & \#Params(M) & throughput(samples/s) & consumption(pJ) &Acc(\%)  \\
		\midrule
            &Recurrent SNN\cite{9311226}& N/a & N/a &N/a &50.90 \\
            &Adaptive SRNN\cite{yin2021accurate}& N/a & N/a &N/a &74.20 \\
            &SpikGRU\cite{10.1007/978-3-031-15934-3_30}& 0.28 & N/a &N/a &77.00 \\
            &RadLIF \cite{10.3389/fnins.2022.865897} & N/a &N/a  &N/a &77.40 \\
            &SNN-Delays(2L-2KC)\cite{2023arXiv230617670H}& 1.4 & 2327.54 &7.5E+6 &80.16 \\
            &SNN-Delays(2L-2KC)+ours & 1.4 & \textbf{3517.00} &\textbf{7.2E+6} &80.39 \\
            &SNN-Delays(3L-1KC)\cite{2023arXiv230617670H}& 1.2 & 2179.00 &7.7E+6 &80.29 \\
		  &SNN-Delays(3L-1KC)+ours & 1.2 & 1829.16 &8.9E+6 &80.97 \\
            &SNN-Delays(3L-2KC)\cite{2023arXiv230617670H}& 2.5 & 1717.89 &1.6E+7 &80.69 \\
            &SNN-Delays(3L-2KC)+ours& 2.5 & 1434.48 &1.8E+7 &\textbf{81.02} \\
		\bottomrule
        \multicolumn{6}{l}{\parbox{0.66\linewidth}{\small nL-mKC stands for a model with n hidden layers and kernel count m, where kernel count denotes the number of non-zero elements in the kernel.}} \\
	\end{tabular}
    
 \end{table*}
 
Furthermore, we have also validated our methods on the SHD dataset by using STSC-SNN, RadLIF, and SNN-Delays as the baseline models. As is shown in Table \ref{tab-shd}, our methods have achieved varying degrees of improvement over the baseline models in different aspects including throughput, energy consumption, and classification accuracy. It is worth mentioning that the best results can be obtained if SNN-Delays model is used as the baseline, where the classification accuracy has improved by 0.94\% to reach 96.04\%, establishing a SOTA result for all models on the SHD dataset.

\begin{table*}[htbp]
\caption{Classification accuracy, parameter count, throughput \\ and energy consumption of different models on SHD}
\label{tab-shd}  
    \centering
	\begin{tabular}{cccc cc}
		\toprule	
	    &Methods & \#Params(M) & throughput(samples/s) & consumption(pJ) &Acc(\%)  \\
		\midrule
            &Cuba-LIF\cite{10.1007/978-3-031-15934-3_30} & 0.14 & N/a&N/a&87.80 \\
            &Adaptive SRNN\cite{yin2021accurate} & N/a & N/a&N/a&90.40 \\
            &STSC-SNN$^*$ \cite{stcc-SNN} & 2.1 & 30419.62 &9.1E+6&91.96 \\
            &STSC-SNN+ours & 2.1 & \textbf{37216.90} &9.1E+6 &92.05 \\
            &Adaptive Delays\cite{10.3389/fnins.2023.1275944} & {0.1} &N/a &N/a &92.45 \\
            &RadLIF$^*$ \cite{10.3389/fnins.2022.865897} & 3.9 & 6384.20 &1.6E+7 &94.32 \\
            &RadLIF+ours & 3.9 & 6624.20 &1.0E+7 &94.18 \\
            &SNN-Delays\cite{2023arXiv230617670H} & 0.2 & 5947.21 &2.1E+6  &95.07 \\
	      &{SNN-Delays+ours} & {0.2} & {6383.48} &\textbf{2.0E+6} &\textbf{96.04} \\
		\bottomrule
	\end{tabular}
 \end{table*}
 \begin{table*}[htbp]
\caption{Classification accuracy, parameter count, throughput \\ and energy consumption of different models on GSC}
\label{tab-gsc}  
    \centering
	\begin{tabular}{cccc cc}
		\toprule	
	    &Methods & \#Params(M) & throughput(samples/s) & consumption(pJ) &Acc(\%)  \\
		\midrule
            &RadLIF \cite{10.3389/fnins.2022.865897} & N/a &N/a  &N/a &77.4 \\
            &SNN-Delays(2L-2KC)\cite{2023arXiv230617670H}& 1.4 & 4956.98 &9.9E+6 &95.00 \\
            &SNN-Delays(2L-2KC)+ours & 1.4 & \textbf{5287.61} &\textbf{9.3E+6} &95.25 \\ 
            &SNN-Delays(3L-1KC)\cite{2023arXiv230617670H} & 1.2 & 3095.54 &1.1E+7 &95.29 \\
            &SNN-Delays(3L-1KC)+ours & 1.2 & 2682.41 &1.2E+7 &95.56 \\
            &SNN-Delays(3L-2KC)\cite{2023arXiv230617670H}& 2.5 & 2741.84 &2.3E+7 &95.35 \\
            &{SNN-Delays(3L-2KC)+ours}& {2.5} & {3303.39} &{2.2E+7} &\textbf{95.62} \\
		\bottomrule
	\end{tabular}
 \end{table*}
 
To further validate the effectiveness of our methods, we have conducted some experiments on non-spike dataset GSC. The experimental results are shown in Table \ref{tab-gsc}. From this table, we can find that our methods consistently provide improvements over the baseline model for one or more aspects of throughput, energy consumption, and classification accuracy, which implies that the proposed methods of this paper are effective on non-spike dataset. In all, we have provided an effective method for different models on both spike and non-spike datasets.

\subsection{Ablation Experiments}
\begin{table}[htbp]
	\centering
	\caption{Ablation Experiments \\ on SSC}
	\label{ab-ssc}  
	\begin{tabular}{ccccc}
	\toprule
		NAR & TR-o &  TR-no&Pool &Acc(\%)  \\
	\midrule
         &  &  & & 80.69 \\
        \checkmark&  &  & &80.99 \\
        \checkmark&  &  & \checkmark&  80.80 \\
        \checkmark&  \checkmark&  & & \textbf{81.02} \\
         \checkmark&  &\checkmark  & & 80.81 \\
         \checkmark&\checkmark &\checkmark  & & 80.85 \\
	\bottomrule
	\end{tabular}
\end{table}
\begin{table}[htbp]
	\centering
	\caption{Ablation Experiments \\ on SHD}
	\label{ab-shd}  
	\begin{tabular}{ccccc}
        \toprule
		NAR & TR-o &  TR-no&Pool &Acc(\%)  \\
        \midrule
         &  &  & & 95.07 \\
        \checkmark&  &  & & 95.75 \\
        \checkmark&  &  & \checkmark&  95.33 \\
        \checkmark&  \checkmark&  & & 95.87 \\
         \checkmark&  &\checkmark  & & \textbf{96.04} \\
        \bottomrule	
	\end{tabular}
\end{table}
\begin{table}[htbp]
	\centering
	\caption{Ablation Experiments \\ on GSC}
	\label{ab-gsc}  
	\begin{tabular}{ccccc}
	\toprule	
		NAR & TR-o &  TR-no&Pool &Acc(\%)  \\
	\midrule
         &  &  & & 95.35 \\
        \checkmark&  &  & & 95.56 \\
        \checkmark&  &  & \checkmark&95.25 \\
        \checkmark&  \checkmark&  & & 95.27 \\
         \checkmark&  &\checkmark  & & \textbf{95.62} \\
         \checkmark&\checkmark  &\checkmark  & & 95.32 \\
	\bottomrule
	\end{tabular}
\end{table}
We have conducted some ablation experiments on the SSC, SHD and GSC datasets for our proposed TR and NAR methods, as are shown in Table \ref{ab-ssc}, Table \ref{ab-shd} and Table \ref{ab-gsc}. 
TR-o uses a group-within time length of 2, with a stride of 1, while TR-no uses a group-within time length of 3 or 2, with a stride of 1. 
From Table \ref{ab-ssc}, it can be seen that we can achieve the best classification accuracy 81.02\% if we use NAR and TR-o.
From Table \ref{ab-shd}, it can be seen that we can achieve the best classification accuracy 96.04\% if we use NAR and TR-no.
From Table \ref{ab-gsc}, it can be seen that we can achieve the best classification accuracy 95.62\% if we use NAR and TR-no.
As a result, we can conclude that the proposed NAR and TR method can demonstrate beneficial effects on three datasets. We have also tested using 1D max pooling instead of our TR, and our TR performed better than 1D max pooling under the same settings.

\section{Conclusion}
\label{sec:conclusion}
In order to learn the information of input data at different temporal scales and apply the residual connection to data of varying time lengths, we propose TR and NAR methods in this paper.
TR groups the input data along time dimensions and then uses logical OR or MAX operations to reconstruct the frequency information of multiple time points within each group into the frequency information of a new single time point, which
is divided into TR with overlap and TR without overlap. 
By using TR, we enable the model to learn information about the input data on a smaller time scale at the shallow layers and learn it on a larger time scale at the deep layers. As a result, the models can process data faster than baselines and can learn information at different temporal scales of the speech data.
NAR allows the different time lengths data to perform residual connection, enabling the models with data of unequal lengths in the temporal dimension to benefit from this structure. 
By using the SNN-Delays model as baseline, we have achieved the desired results on the spiking datasets SSC, SHD, and the non-spiking dataset GSC through numerous experiments. On the dataset SSC, we have achieved the SOTA result 81.02\% for the test classification accuracy of all SNN models. And we have obtained the SOTA result 96.04\% on the dataset SHD for the classification accuracy of all models. On the non-spiking dataset GSC, our methods also have brought improvements to the baseline model. Meanwhile, our methods have also increased the models' throughput and reduced their energy consumption.





\ifCLASSOPTIONcaptionsoff
  \newpage
\fi

\bibliographystyle{IEEEtran}
\bibliography{ref}

\begin{thebibliography}{10}
\providecommand{\url}[1]{#1}
\csname url@samestyle\endcsname
\providecommand{\newblock}{\relax}
\providecommand{\bibinfo}[2]{#2}
\providecommand{\BIBentrySTDinterwordspacing}{\spaceskip=0pt\relax}
\providecommand{\BIBentryALTinterwordstretchfactor}{4}
\providecommand{\BIBentryALTinterwordspacing}{\spaceskip=\fontdimen2\font plus
\BIBentryALTinterwordstretchfactor\fontdimen3\font minus \fontdimen4\font\relax}
\providecommand{\BIBforeignlanguage}[2]{{%
\expandafter\ifx\csname l@#1\endcsname\relax
\typeout{** WARNING: IEEEtran.bst: No hyphenation pattern has been}%
\typeout{** loaded for the language `#1'. Using the pattern for}%
\typeout{** the default language instead.}%
\else
\language=\csname l@#1\endcsname
\fi
#2}}
\providecommand{\BIBdecl}{\relax}
\BIBdecl

\bibitem{doi:10.1142/S0129065709002002}
S.~Ghosh-dastidar and H.~Adeli, ``Spiking neural networks,'' \emph{International Journal of Neural Systems}, vol.~19, no.~4, pp. 295--308, 2009.

\bibitem{Xu_2023_CVPR}
Q.~Xu, Y.~Li, J.~Shen, J.~K. Liu, H.~Tang, and G.~Pan, ``Constructing deep spiking neural networks from artificial neural networks with knowledge distillation,'' in \emph{Proceedings of the IEEE/CVF Conference on Computer Vision and Pattern Recognition(CVPR)}, Jun. 2023, pp. 7886--7895.

\bibitem{MAASS19971659}
W.~Maass, ``Networks of spiking neurons: The third generation of neural network models,'' \emph{Neural Networks}, vol.~10, no.~9, pp. 1659--1671, 1997.

\bibitem{9597475}
Y.~Hu, H.~Tang, and G.~Pan, ``Spiking deep residual networks,'' \emph{IEEE Transactions on Neural Networks and Learning Systems}, vol.~34, no.~8, pp. 5200--5205, 2023.

\bibitem{10003253}
Q.~Xu, Y.~Li, J.~Shen, P.~Zhang, J.~K. Liu, H.~Tang, and G.~Pan, ``Hierarchical spiking-based model for efficient image classification with enhanced feature extraction and encoding,'' \emph{IEEE Transactions on Neural Networks and Learning Systems}, vol.~35, no.~7, pp. 9277--9285, 2024.

\bibitem{wu2022brain}
Y.~Wu, R.~Zhao, J.~Zhu, F.~Chen, M.~Xu, G.~Li, S.~Song, L.~Deng, G.~Wang, H.~Zheng \emph{et~al.}, ``Brain-inspired global-local learning incorporated with neuromorphic computing,'' \emph{Nature Communications}, vol.~13, no.~1, pp. 65--78, 2022.

\bibitem{10168967}
T.~Zhang, Q.~Wang, and B.~Xu, ``Self-lateral propagation elevates synaptic modifications in spiking neural networks for the efficient spatial and temporal classification,'' \emph{IEEE Transactions on Neural Networks and Learning Systems}, vol.~35, no.~11, pp. 15\,359--15\,371, 2024.

\bibitem{Liao_Liu_Zheng_Pan_2024}
Z.~Liao, Y.~Liu, Q.~Zheng, and G.~Pan, ``Spiking nerf: Representing the real-world geometry by a discontinuous representation,'' in \emph{Proceedings of the AAAI Conference on Artificial Intelligence (AAAI)}, vol.~38, no.~12, Mar. 2024, pp. 13\,790--13\,798.

\bibitem{bu2023optimal}
T.~Bu, W.~Fang, J.~Ding, P.~Dai, Z.~Yu, and T.~Huang, ``Optimal ann-snn conversion for high-accuracy and ultra-low-latency spiking neural networks,'' in \emph{International Conference on Learning Representations(ICLR)}, Apr. 2022, pp. 1--20.

\bibitem{pmlr-v202-wang23j}
Z.~Wang, R.~Jiang, S.~Lian, R.~Yan, and H.~Tang, ``Adaptive smoothing gradient learning for spiking neural networks,'' in \emph{Proceedings of the 40th International Conference on Machine Learning(ICML)}, Jul. 2024, pp. 35\,798--35\,816.

\bibitem{shen2024rethinking}
H.~Shen, Q.~Zheng, H.~Wang, and G.~Pan, ``Rethinking the membrane dynamics and optimization objectives of spiking neural networks,'' in \emph{The Thirty-eighth Annual Conference on Neural Information Processing Systems(NIPS)}, Dec. 2024, pp. 1--24.

\bibitem{9540752}
M.~Zhang, J.~Wang, J.~Wu, A.~Belatreche, B.~Amornpaisannon, Z.~Zhang, V.~P.~K. Miriyala, H.~Qu, Y.~Chua, T.~E. Carlson, and H.~Li, ``Rectified linear postsynaptic potential function for backpropagation in deep spiking neural networks,'' \emph{IEEE Transactions on Neural Networks and Learning Systems}, vol.~33, no.~5, pp. 1947--1958, 2022.

\bibitem{Wang_Zhang_Han_Wang_Zhang_Xu_2023}
Q.~Wang, T.~Zhang, M.~Han, Y.~Wang, D.~Zhang, and B.~Xu, ``Complex dynamic neurons improved spiking transformer network for efficient automatic speech recognition,'' in \emph{Proceedings of the AAAI Conference on Artificial Intelligence (AAAI)}, vol.~37, no.~1, Jun. 2023, pp. 102--109.

\bibitem{10254579}
Y.~Hu, Q.~Zheng, X.~Jiang, and G.~Pan, ``Fast-snn: Fast spiking neural network by converting quantized ann,'' \emph{IEEE Transactions on Pattern Analysis and Machine Intelligence}, vol.~45, no.~12, pp. 14\,546--14\,562, 2023.

\bibitem{MA2025106765}
Y.~Ma, H.~Wang, H.~Shen, S.~Duan, and S.~Wen, ``Analog spiking u-net integrating cbam\&vit for medical image segmentation,'' \emph{Neural Networks}, vol. 181, pp. 106\,765--1--12, 2025.

\bibitem{10032591}
M.~Yao, G.~Zhao, H.~Zhang, Y.~Hu, L.~Deng, Y.~Tian, B.~Xu, and G.~Li, ``Attention spiking neural networks,'' \emph{IEEE Transactions on Pattern Analysis and Machine Intelligence}, vol.~45, no.~8, pp. 9393--9410, 2023.

\bibitem{ma2025neuromoco}
Y.~Ma, H.~Wang, H.~Shen, X.~Chen, S.~Duan, and S.~Wen, ``Neuromoco: a neuromorphic momentum contrast learning method for spiking neural networks,'' \emph{Applied Intelligence}, vol.~55, no.~2, pp. 1--13, 2025.

\bibitem{9646435}
J.~Shen, Y.~Zhao, J.~K. Liu, and Y.~Wang, ``Hybridsnn: Combining bio-machine strengths by boosting adaptive spiking neural networks,'' \emph{IEEE Transactions on Neural Networks and Learning Systems}, vol.~34, no.~9, pp. 5841--5855, 2023.

\bibitem{stcc-SNN}
C.~Yu, Z.~Gu, D.~Li, G.~Wang, A.~Wang, and E.~Li, ``Stsc-snn: Spatio-temporal synaptic connection with temporal convolution and attention for spiking neural networks,'' \emph{Frontiers in Neuroscience}, vol.~16, pp. 1\,079\,357--1--15, 2022.

\bibitem{10.3389/fnins.2023.1275944}
P.~Sun, Y.~Chua, P.~Devos, and D.~Botteldooren, ``Learnable axonal delay in spiking neural networks improves spoken word recognition,'' \emph{Frontiers in Neuroscience}, vol.~17, pp. 1\,275\,944--1--12, 2023.

\bibitem{2023arXiv230617670H}
I.~Hammouamri, I.~Khalfaoui-Hassani, and T.~Masquelier, ``Learning delays in spiking neural networks using dilated convolutions with learnable spacings,'' in \emph{The Twelfth International Conference on Learning Representations(ICLR)}, May. 2024, pp. 1--14.

\bibitem{9311226}
B.~Cramer, Y.~Stradmann, J.~Schemmel, and F.~Zenke, ``The heidelberg spiking data sets for the systematic evaluation of spiking neural networks,'' \emph{IEEE Transactions on Neural Networks and Learning Systems}, vol.~33, no.~7, pp. 2744--2757, 2020.

\bibitem{10.1162/neco.2009.11-08-901}
F.~Ponulak and A.~Kasiński, ``{Supervised Learning in Spiking Neural Networks with ReSuMe: Sequence Learning, Classification, and Spike Shifting},'' \emph{Neural Computation}, vol.~22, no.~2, pp. 467--510, Feb. 2010.

\bibitem{yin2021accurate}
B.~Yin, F.~Corradi, and S.~M. Boht{\'e}, ``Accurate and efficient time-domain classification with adaptive spiking recurrent neural networks,'' \emph{Nature Machine Intelligence}, vol.~3, no.~10, pp. 905--913, 2021.

\bibitem{10.3389/fnins.2022.865897}
A.~Bittar and P.~N. Garner, ``A surrogate gradient spiking baseline for speech command recognition,'' \emph{Frontiers in Neuroscience}, vol.~16, pp. 865\,897--1--18, 2022.

\bibitem{10.1007/978-3-031-15934-3_30}
M.~Dampfhoffer, T.~Mesquida, A.~Valentian, and L.~Anghel, ``Investigating current-based and gating approaches for accurate and energy-efficient spiking recurrent neural networks,'' in \emph{Artificial Neural Networks and Machine Learning -- ICANN 2022}, Sep. 2022, pp. 359--370.

\bibitem{gerstner2002spiking}
W.~Gerstner and W.~M. Kistler, \emph{Spiking neuron models: Single neurons, populations, plasticity}.\hskip 1em plus 0.5em minus 0.4em\relax Cambridge university press, 2002.

\bibitem{Yao_2021_ICCV}
M.~Yao, H.~Gao, G.~Zhao, D.~Wang, Y.~Lin, Z.~Yang, and G.~Li, ``Temporal-wise attention spiking neural networks for event streams classification,'' in \emph{Proceedings of the IEEE/CVF International Conference on Computer Vision (ICCV)}, Oct. 2021, pp. 10\,221--10\,230.

\bibitem{BORNKESSELSCHLESEWSKY2015142}
I.~Bornkessel-Schlesewsky, M.~Schlesewsky, S.~L. Small, and J.~P. Rauschecker, ``Neurobiological roots of language in primate audition: common computational properties,'' \emph{Trends in Cognitive Sciences}, vol.~19, no.~3, pp. 142--150, 2015.

\bibitem{deHeer6539}
W.~A. de~Heer, A.~G. Huth, T.~L. Griffiths, J.~L. Gallant, and F.~E. Theunissen, ``The hierarchical cortical organization of human speech processing,'' \emph{Journal of Neuroscience}, vol.~37, no.~27, pp. 6539--6557, 2017.

\bibitem{8047110}
F.~Rong, ``Audio classification method based on machine learning,'' in \emph{2016 International Conference on Intelligent Transportation, Big Data \& Smart City (ICITBS)}, Dec. 2016, pp. 81--84.

\bibitem{10095023}
S.~Ghosh, A.~Seth, S.~Umesh, and D.~Manocha, ``Mast: Multiscale audio spectrogram transformers,'' in \emph{ICASSP 2023 - 2023 IEEE International Conference on Acoustics, Speech and Signal Processing (ICASSP)}, Jun. 2023, pp. 1--5.

\bibitem{burkitt2006review}
A.~N. Burkitt, ``A review of the integrate-and-fire neuron model: I. homogeneous synaptic input,'' \emph{Biological cybernetics}, vol.~95, pp. 1--19, 2006.

\bibitem{doi:10.1126/sciadv.adi1480}
W.~Fang, Y.~Chen, J.~Ding, Z.~Yu, T.~Masquelier, D.~Chen, L.~Huang, H.~Zhou, G.~Li, and Y.~Tian, ``Spikingjelly: An open-source machine learning infrastructure platform for spike-based intelligence,'' \emph{Science Advances}, vol.~9, no.~40, pp. eadi1480--1--18, 2023.

\bibitem{khalfaouihassani:hal-04057309}
I.~Khalfaoui-Hassani, T.~Pellegrini, and T.~Masquelier, ``{Dilated convolution with learnable spacings},'' in \emph{{11th International Conference on Learning Representations (ICLR 2023)}}, May. 2023, pp. 1--29.

\bibitem{MAASS199926}
W.~Maass and M.~Schmitt, ``On the complexity of learning for spiking neurons with temporal coding,'' \emph{Information and Computation}, vol. 153, no.~1, pp. 26--46, 1999.

\bibitem{Liu_2021_ICCV}
Z.~Liu, Y.~Lin, Y.~Cao, H.~Hu, Y.~Wei, Z.~Zhang, S.~Lin, and B.~Guo, ``Swin transformer: Hierarchical vision transformer using shifted windows,'' in \emph{Proceedings of the IEEE/CVF International Conference on Computer Vision (ICCV)}, Oct. 2021, pp. 10\,012--10\,022.

\bibitem{9746312}
K.~Chen, X.~Du, B.~Zhu, Z.~Ma, T.~Berg-Kirkpatrick, and S.~Dubnov, ``Hts-at: A hierarchical token-semantic audio transformer for sound classification and detection,'' in \emph{ICASSP 2022 - 2022 IEEE International Conference on Acoustics, Speech and Signal Processing (ICASSP)}, May. 2022, pp. 646--650.

\bibitem{He_2016_CVPR}
K.~He, X.~Zhang, S.~Ren, and J.~Sun, ``Deep residual learning for image recognition,'' in \emph{Proceedings of the IEEE Conference on Computer Vision and Pattern Recognition (CVPR)}, Jun. 2016, pp. 770--778.

\bibitem{SHEN2024120136}
H.~Shen, H.~Wang, Y.~Ma, L.~Li, S.~Duan, and S.~Wen, ``Multi-lra: Multi logical residual architecture for spiking neural networks,'' \emph{Information Sciences}, vol. 660, pp. 120\,136--1--13, 2024.

\bibitem{10486935}
Y.~Hu, Q.~Zheng, and G.~Pan, ``Bitsnns: Revisiting energy-efficient spiking neural networks,'' \emph{IEEE Transactions on Cognitive and Developmental Systems}, vol.~16, no.~5, pp. 1736--1747, 2024.

\bibitem{ladd2008intonational}
D.~R. Ladd, \emph{Intonational phonology}.\hskip 1em plus 0.5em minus 0.4em\relax Cambridge University Press, 2008.

\bibitem{6757323}
M.~Horowitz, ``1.1 computing's energy problem (and what we can do about it),'' in \emph{2014 IEEE International Solid-State Circuits Conference Digest of Technical Papers (ISSCC)}, Feb. 2014, pp. 10--14.

\end{thebibliography}

\end{document}